Title: Computer-assisted vaccine design

Hao Zhou, Ramdas S. Pophale, and Michael W. Deem
Departments of Bioengineering and Physics & Astronomy
Rice University
Houston, TX, USA

Abstract

We define a new parameter to quantify the antigenic distance between two H3N2 influenza strains: we use this parameter to measure antigenic distance between circulating H3N2 strains and the closest vaccine component of the influenza vaccine. For the data between 1971 and 2004, the measure of antigenic distance correlates better with efficacy in humans of the H3N2 influenza A annual vaccine than do current state of the art measures of antigenic distance such as phylogenetic sequence analysis or ferret antisera inhibition assays. We suggest that this measure of antigenic distance could be used to guide the design of the annual flu vaccine. We combine the measure of antigenic distance with a multiple-strain avian influenza transmission model to study the threat of simultaneous introduction of multiple avian influenza strains. For H3N2 influenza, the model is validated against observed viral fixation rates and epidemic progression rates from the World Health Organization FluNet – Global Influenza Surveillance Network. We find that a multiple-component avian influenza vaccine is helpful to control a simultaneous multiple introduction of bird-flu strains. We introduce Population at Risk (PaR) to quantify the risk of a flu pandemic, and calculate by this metric the improvement that a multiple vaccine offers.

# Computer-assisted vaccine design

*Introduction*

Circulating influenza virus uses the ability to change its surface proteins, along with its high transmission rate, to flummox the adaptive immune response of the host. The random accumulation of mutations in the hemagglutinin (HA) and neuraminidase (NA) epitopes, the regions on surface of the viral proteins that are recognized by host antibodies, pose a formidable challenge to the design of an effective annual flu vaccine. Under current practice, the World Health Organization (WHO), and the respective government health agencies, rely on historical experience and phylogenetic analysis of HA and NA protein sequences from the circulating human strains to decide upon the components in the annual influenza vaccine [Munoz et. al., 2003]. Every year, the concerned authorities make a projection about the circulating influenza strains for the coming flu season. The flu vaccine currently contains three strains that are as similar as possible to those strains that are predicted to be the most prominent. At present, the three strains in the vaccine are one H3N2 A, one H1N1 A, and one influenza B component. Historically, the vaccine efficacy has seldom reached the 100% mark. Over the years, it has hovered between 30-60% against influenza-like illnesses. In fact, due to a phenomenon known as the 'original antigenic sin' [Davenport et. al., 1953; Fazekas et. al., 1966; Deem et. al., 2003], the vaccine efficacy has even been negative at times. If original antigenic sin is operative, the host antibodies that are produced in response to a viral strain tend to suppress creation of new and different antibodies in response to a different viral strain. Whether such a phenomenon takes place or whether the vaccine is effective depends to a great extent on how similar the vaccine component strains are to the circulating viral strains. The methods employed to calculate such an antigenic distance draw heavily from the ferret antisera hemagglutinin inhibition assays. It has been assumed that the antigenic distance thus obtained from ferrets correlates well with the efficacy of the influenza vaccine in humans. However, to our knowledge there is limited evidence in the literature of such correlations.

We introduce the reader to a new and effective way of measuring the antigenic distance between different strains. We define a quantity, $p_{epitope}$, that measures the difference between the dominant epitope regions of the hemagglutinin proteins of any two H3N2 influenza viruses in general. The term $p_{epitope}$ can be used to measure difference between the dominant epitope of the H3N2 vaccine component and the dominant circulating viral strain. A dominant epitope is one that elicits the most significant response from the adaptive immune system for a particular strain in a particular year [Fitch et. al., 1991; Bush et. al., 1999; Fitch et. al., 2000; Plotkin et. al., 2003]. We show that for data spanning last 35 years, the $p_{epitope}$ measure of antigenic distance correlates better with the efficacy studies in humans of the influenza vaccine than do the current measures of antigenic distance, even those derived from ferret animal model studies. The $p_{epitope}$ measurement can also be used to measure difference between two circulating viral strains. The $p_{epitope}$ measurement can give us an idea of antigenic variability of circulating influenza strains.

*Methods*

We have developed a statistical mechanics based theory to model the response of an immune system, free of immunoscenescence, to disease and vaccination. We compare this theory to experimental studies of vaccine efficacies for 18-64 year old subjects over the past 35 years, when the H3N2 subtype of influenza A was the dominant strain. H3N2 is the most common stain and has caused significant morbidity and mortality [Macken et. al., 2001]. As is the norm, we focused on the five epitopes of the hemagglutinin protein. We used the generalized NK model to calculate the affinity constants quantifying the immune response following exposure to an antigen after vaccination. The model takes into account three types of interactions within an antibody. These include interactions within subdomains, interactions between different subdomains, and interaction between an antibody and an antigen [Jun et. al., 2005]. The binding constant is given as $K = \exp(a-b<U>)$, where U is the energy function of an antibody [Deem et. al. 2003], $a = -18.56$, and $b = 1.67$. The values of the constants are determined after comparing the dynamics of the model with experimental results [Deem et. al., 2003]. In order to capture the antigenic distance between the vaccine strain and the circulating strain, we use $p_{epitope}$ as an order parameter in our model. It represents the fraction of amino acids that differ between the dominant epitope regions of the two strains.

$$p_{epitope} = \frac{\text{number of amino acid differences in the dominant epitope}}{\text{total number of amino acids in the dominant epitope}} \qquad (1)$$

In order to build the model we needed the identity and sequence of the dominant epitope in both the candidate vaccine and the circulating strain. This definition of the five epitopes in the H3N2 hemagglutinin protein was taken from reference [Macken et. al., 2001]. Our model assumes the following correlation between the vaccine efficacy, E, and the binding constants. $E = \alpha \ln[K_{secondary}(p_{epitope}) / K_{primary}]$, where the constant α is selected such that a perfect match between the vaccine and the circulating strain corresponds to the average historical vaccine efficacy of 45% [Gupta et. al., 2006]. $K_{primary}$ is the binding constant corresponding to the primary immune response, and $K_{secondary}$ is the binding constant corresponding to the secondary immune response that follows vaccination. Other than the constant, α, no parameter was fit to data, making the model predictive. For example, the point where vaccine efficacy becomes zero is independent of the value of α.

*Results*

Table 1 compares vaccine efficacy values from experimental studies [Smith et. al., 1979; Clements et. al., 1986; Keitel et. al., 1988; Edwards et. al., 1994; Nichol et. al., 1995;

Keitel et. al., 1997; Campbell et. al., 1997; Grotto et. al., 1998; Bridges et. al., 2000; Mixeu et. al., 2002; Millot et. al., 2002; Kawai et. al., 2003; Lester et. al., 2003; Dolan et. al., 2004] and predictions from our theory as a function of the $p_{epitope}$. In literature, vaccine efficacy is defined as

$$\text{Efficiency} = \frac{u - v}{u}, \tag{2}$$

where u and v are the influenza-like illness rates in unvaccinated and vaccinated individuals, respectively. The epidemiological estimates of u and v contain some noise; however, these are the best estimates there are of influenza vaccine efficacy in humans. Our model demonstrates the effectiveness of using $p_{epitope}$ as a measure of antigenic drift between the circulating strain and the vaccine strain. Crystallographic data and immunoassays have shown that only the epitope regions of the viral proteins are significantly involved in the recognition of the antigen via the antibodies [Air et. al., 1985]. Our definition of $p_{epitope}$ follows from this observation. When $p_{epitope}$ value in the dominant epitope is greater than 0.19 according to experimental records or is greater than 0.22 according to our theory, the efficacy of the influenza vaccine drops to the negative territory [Table 1 and Figure 1a] indicating that while designing the vaccine, this regime needs to be avoided. As an example, during the 1997/98 influenza season in the Northern hemisphere, when the Sydney/5/97 strain was widespread, the value of $p_{epitope}$ was 0.238 resulting in an efficacy of -17% [Bridges et. al., 2000]. Only one data point falls outside our theoretical predications, that for the 1989/90 epidemic [Nguyen-Van-Tran et. al., 2003]. During that year there were likely multiple strains circulating, including the strains of influenza B [Edwards et. al., 1994; Ikonen et. al., 1996].

The WHO uses as a first approximation an alternative definition of the antigenic drift. It considers the sequence difference of the entire hemagglutinin protein of the circulating and viral strains.

$$p_{sequence} = \frac{\text{number of amino acid differences in the sequence}}{\text{total number of amino acids in the sequence}} \tag{3}$$

The correlation with the experimentally observed efficacy is not as positive when using the $p_{sequence}$ definition for the antigenic distance. This conclusion is seen from Table 1 and Figure 1b, and it follows from the fact that many domains in the protein are either inaccessible to the human antibodies or not recognizable with high probability by them. Immune recognition of the epitope regions results in the vaccine efficacy being more correlated with the $p_{epitope}$ measure of antigenic distance in Eq. 1 than with the approximate measure of distance in Eq. 3.

The gold standard measure of antigenic drift is that derived from ferret antisera [Smith et. al., 1999; Lee et. al., 2004]. Comparison of the vaccine efficacy to this measure of antigenic drift also shows less than ideal success. From Table 1 and Figure 1c, the hemagglutinin assays in ferret antisera are unable to capture a significant amount of human efficacy information. For example, an antigenic distance of zero according to ferret antisera experiments does not mean that the two strains are identical. As an example, for the 1996/97 season, the antigenic distance derived from ferret antisera between the vaccine strain of A/Nanchang/933/95 and the circulating strain of A/Wuhan/359/95 was zero, whereas $p_{epitope}$ value was 0.095. The corresponding vaccine efficacy in the Northern and Southern hemispheres was 28% [Millot et. al., 2002] and 11% [Mixeu et. al., 2002] respectively. When compared to the average efficacy corresponding to a perfect match between the vaccine strain and the circulating strain, which is 45%, these efficacy values are clearly much lower. Thus, ferret derived distances will occasionally misidentify antigenically distant strains as antigenically identical.

*Discussion*

Design of the influenza vaccine is always a race against time. Currently, under the supervision of the WHO and the national health agencies, in the Northern hemisphere, the components of the annual influenza vaccine are determined between February and April. The mass production of the vaccine is then carried out by growing the virus in hen's eggs. After regulatory tests in mid-July, the vaccine is distributed in September. Data are collected to determine the vaccine efficacy starting in October and continuing through the Winter flu season. By January a good measure of the effectiveness of the season's vaccine is obtained. Choice of the vaccine strain is contingent up on various biological and manufacturing constraints. For example, among the egg-cultured strains, the availability of high growth strains is an additional criterion. The organizations in charge have to weigh in all these constraints before they decide on the best match for the anticipated circulating strain for the following flu season.

The $p_{epitope}$ measure of antigenic distance can influence and contribute to the vaccine design process in two ways. First, $p_{epitope}$ can help to identify the strains to be included in the annual flu vaccine. For every season, given a list of strains and their probabilities of outbreak, the value of $p_{epitope}$ can help define the weighted distance of a particular vaccine strain from the circulating strains. This procedure will enable us to identify the vaccine strain closest to the strains causing a possible outbreak. Alternately, $p_{epitope}$ can also help identify the 'like' strains and to quantify 'likeness.' Various manufacturing constraints in the vaccine production process mean it is often not feasible to grow large quantities of the exact strain that is desired for the annual vaccine. Under such scenario, it is required to choose several strains that are similar to the chosen one. The value of $p_{epitope}$ can help quantify the 'likenesses' of a given strain to the desired strain. We can extend $p_{epitope}$ measure of antigenic distance to other strains of influenza as well. Using information about the epitope regions of other HxNy strains of influenza A or influenza B, we can calculate the value of $p_{epitope}$ to quantify the antigenic distance for these strains of influenza.

It is clear that the immune system response is neither linear nor monotonic in the antigenic distance. As a result, the original antigenic sin leading to a negative vaccine efficacy exists only in an intermediate regime of the antigenic distance. If the vaccine falls in this regime, however, the vaccinated individual appears to be more susceptible to influenza-like illnesses compared to an unvaccinated individual. Over the past 33 years, this phenomenon seems to have taken place 26% of the time according to the H3N2 epidemiological data as indicated by 5 of the total 19 data points that are negative in table 1 and figure 1.b. The negative points are not necessarily a result of experimental errors; rather they may have their roots in the original antigenic sin phenomenon where the immune system is unable to distinguish between the new and the old viral strains. It is desired that the regime corresponding to the original antigenic sin is avoided, not only for the obvious immunological consequences, but also for the negative impact it creates against the acceptance of public health policy. Our theory rightfully captures the underlying physics of immune response, and corroborates the experimental findings. Our theory and $p_{epitope}$ measure of antigenic distance can be used to ensure that vaccines chosen would not fall in the original antigenic sin region against expected strains. Saying this more conservatively, our theory can be used to ensure that the vaccine strains chosen are antigenically close enough to the expected strains so that the vaccine efficacy is expected to be positive.

It appears that $p_{epitope}$ is a measure of the antigenic distance between viral strains that correlates more strongly with vaccine efficacy than is the sequence analysis or the ferret antisera. The population at large may benefit if the health authorities incorporate $p_{epitope}$ in the prediction and design of the influenza vaccine in addition to (or even instead of) the current practices. We now present an example to show how our theory can assist shaping public health policies. We examine the 2004/2005 flu season in the Northern hemisphere. With the aid of $p_{epitope}$ we can stipulate a priori the extent of protection that a particular vaccine strain provides against the circulating strain for a particular season. The comparison among various candidates would provide us selection criteria for their inclusion in the annual flu vaccine. During the 2003/2004 flu epidemic, A/Fujian/411/2002 strain was predominant, and it was expected that this strain would dominant in 2004/2005 as well. In order to counter this strain, the Advisory Council of the FDA recommended using A/Wyoming/2003 as the H3N2 component of the 2004/2005 vaccine, since according to the prevalent measures, this strain and A/Fujian/411/2002 were found to be 'antigenically equivalent' [Harper et. al., 2004]. Our calculations, however, yielded a $p_{epitope}$ value of 0.095 for the pair in question, suggesting that the two strains were not antigenically equivalent, and the predicted efficacy is 20% (figure 1.a). Also, in 2004/2005, A/California/7/2004 strain was in circulation along with significant amount of influenza B. The $p_{epitope}$ value for the pair of A/California/7/2004 and the A/Wyoming/2003 strain was 0.286. According to our theory the vaccine would not have provided a positive protection from the A/California/7/2004 strain. The vaccine efficacy that year was 9.2% [Chan et. al., 2008], which is roughly the average of the efficacies (0% and 20%) predicted by $p_{epitope}$ theory. We note that another of the WHO approved candidates for that year was A/Kumamoto/102/02 (ISDN38180) [WHO, 2004], and we found $p_{epitope}$ value to be zero versus A/Fujian/411/2002. According to our theory, a vaccine containing the Kumamoto strain would have provided a better protection against the Fujian strain that did the Wyoming vaccine strain.

*Suggestions to improve predictability of vaccine efficacy*

At the heart of our approach lies the identification of the dominant hemagglutinin epitope recognized in humans. The identity of the dominant epitope for humans is currently not measured. We, thus, define and postulate the dominant epitope for humans for a certain strain and for a certain season as the epitope that undergoes the largest fractional change in the sequence of amino acids in comparison to the vaccine strain. A measurement of the epitope that is dominant in humans for various circulating strains and vaccines should help to improve the predictive power of our approach even more, as we can then use the measured dominant epitope to calculate $p_{epitope}$ instead of using our postulated dominant epitope. It is important to continue measurement and sequence analysis of the prominent circulating strains in the flu season. These data will help to validate and perhaps better calibrate the $p_{epitope}$ measure of antigenic distance. Another important piece of the puzzle is epidemiological study to relate vaccine efficacy to the antigenic drift. We believe that use of the $p_{epitope}$ measure of antigenic distance will enable the health authorities to predict the severity of the yearly flu season, to design better vaccines for the same, and help to manage the health resources for this period.

In more general terms, the results presented here have implications for the fight against diseases that stem from rapidly mutating viruses and that are managed through antibody responses. The $p_{epitope}$ measure of antigenic distance can predict efficacies for vaccine strains against multiple circulating strains, and it could then be used towards redesigning of the vaccine in terms of frequency and composition. One such disease that has been threatening to turn into a pandemic of late is the avian influenza, or bird flu disease. In the next section, we present a multiple-strain transmission model for avian influenza that would enable one to manage the risk from such an outbreak. We use $p_{epitope}$ to evaluate the effectiveness of a multiple-component vaccine to counter such threat.

**Vaccination strategies and risk management of flu pandemic**

Since its first appearance (Hong Kong, 1997), H5N1 avian influenza is known to have spread to various parts of the world, including South-East Asian countries, parts of central and Middle-East Asia, Africa, and Europe. This rapid spread has prompted the World Health Organization (WHO) to suggest that "we are closer to a pandemic than at any time since 1968" [WHO report, 2005; Mills et. al., 2006]. Bird flu has been observed in pigs [Cyranoski, 2005], and occurrences of person-to-person transmission have likely been detected [Ungchusak et. al., 2005; Normile, 2006; Yang et. al., 2006]. Efficacy of the vaccine produced against the original Hong Kong strain of H5N1 has been poor against some of the new strains. In addition, the high mutation rates observed in the avian influenza have raised the prospect of simultaneous introduction of multiple strains [Mills et. al., 2006; Ducatez et. al., 2006; Chen et. al., 2006; Capua et. al., 2007], putting a question mark against the efficacy of a single-component vaccine. Although in the event of appearance of multiple strains, eventual emergence of a single dominant strain is likely; the lack of a priori knowledge of which strain will be dominant makes it worthwhile to look into the alternative of a multiple-component vaccine. Here, we use an epidemiological model to study the efficacy and cross-protection of a multiple-component bird flu vaccine. We further extend the concept of $p_{epitope}$ through a stochastic model to manage the risk of a flu pandemic. The strategy is inspired by similar risk management studies in finance, and the population at risk for infection in an epidemic plays the role of the risk variable. Here $p_{epitope}$ represents the difference between circulating strains.

*Multiple-strain introduction transmission model*

As mentioned above, there is evidence, both theoretical and otherwise, that deems introduction of multiple-strains of avian influenza is a distinct possibility. There have been no mathematical models, however, that explore such a scenario to evaluate efficacy of practicable vaccine strategies. Our model uses the concepts of hierarchical scale free network, and virus transmission, and viral evolution to evaluate possible, and worst case, epidemic spreading scenarios and to evaluate efficacy of single or multiple-component vaccine strategies.

*Hierarchical scale free network*

We divide the total human population ($6.7 \times 10^9$) into ($6.7 \times 10^6$) groups. Each group consists of $10^3$ people that exhibit similar health status and social behavior. These groups are distributed over $N_{city}$ cities. The distribution of cities with i groups ($N_{city}(i)$) goes as $N_{city}(i) \propto i^{-2.1}$ [Zipf, 1949; Newman, 2005]. The largest and the smallest city areas in our simulation have $6.2 \times 10^4$ and 400 groups respectively [Newman, 2005]. The total number of cities fluctuates around 4000 [Guimera et. al., 2005].

The cities are connected through a network that extends globally and is defined by the network spawned by the airlines. The distribution of cities with i contacts is given by $N_{city\_contact}(i) \propto i^{-2.0}$. The network connecting groups within a city is defined by the ground transportation network. In each city, the distribution of groups with i contacts is given by $N_{group\_contact}(i) \propto i^{-2.8}$ [Eubank et. al., 2004]. Population size differences between cities mean that movement from one city to another leads to a final averaged distribution over cities that is different for different contact numbers. Any hierarchical structure present in the actual distributions of within- or between-city contacts, which is not detected by the scale-free analyses performed to date, would generally be expected to slow down the spread of an epidemic through the population.

The scale free network is generated in two steps. The first step is to randomly assign a degree to each node based on the given power law distribution. This is achieved by the classical transformation method [Press et al., 2002]. The second step is to connect nodes. This is achieved by connecting nodes with a probability that is proportional to their degree. This scale free network generation method has been used in other scenarios [Albert et al.].

*Virus transmission and evolution*

In our simulation, the viruses are introduced in groups that are chosen randomly. The virus has a latency period of 2 days, which is followed by the infectious period. During the infectious period, the virus can either be killed by the host immune system, or it can be transmitted between different groups globally. During both these periods, the virus can mutate as well as be killed. The term $R_{mutation} = 1.6 \times 10^{-5}$/amino acid/day [Sato et. al., 2006] gives the mutation rate, whereas, the virus killing probability is given by Kill_prob = cKill + cAlpha * max[(cEpiMax – epi), 0]. Here cKill is the intrinsic probability of killing virus, and its baseline value is fitted in our model to be 0.36. A value of 1.0 proved to be optimal for cAlpha. cEpiMax is the largest possible epitope based distance between the vaccine strain and the current viral strain before the vaccine efficacy goes to zero, and its value is 0.19 [Gupta et. al., 2006]. The term epi is a function of the host immune history, and circulating virus strain. It is defined as the smallest distance among the epitope based distances between the current viral strain and the viral strains (50) in the immune history of the infected group. The transmission rate takes into account the seasonal effect [Ferguson et. al., 2003], and is different for transmission within a city (Ffac1×$\tau_0$×(1+0.25×Sin(2πT/360)), Ffac1=1.0) and transmission between cities (Ffac×$\tau_0$×(1+0.25), Ffac=7.0). FCap is the number of different groups that people travelling on a flight come from, and it is set to 100 for our model. The term $\tau_0$ is the intrinsic transmission ability of the virus, and equals 0.07 for the virus introduced on day 1. The mutants, once created, can transmit according to probability cProb. For a baseline parameter, cProb equals 0.26.

*Vaccination Strategies*

As an example, for a very rapid public health policy response, we consider that after 40 days, individuals are vaccinated. We consider single-component as well as multiple-component vaccines. In single-component case, the most dominant strain at day 10 is taken as the vaccine strain, whereas, for multiple-component case, the top 10 strains that are most dominant on day 10 are incorporated into the vaccine. Day 10 is taken as the data collection day by public health authorities. We calculated the efficacy and the cumulative attack rates for both cases. Efficacy is defined by equation 2 above. Thus, efficacy represents the percentile reduction in disease occurrences in people who were vaccinated compared to those who were not [Nichol et. al., 1994; Chow, 2003].

The model's ability to predict an influenza pandemic was tested against the existing epidemiological data (figure 2a). The model successfully predicted the average trend of H3N2 isolates data in FluNet database of WHO from 1995 to 2006 (figure 2b). We also successfully predicted the fixation rate for dominant and non-dominant epitopes of influenza (figure 3) versus those measured [Ferguson et. al., 2003].

*WHO FluNet data analysis*

We acquired the WHO FluNet data for the period of 1995-2006 [WHO FluNet, 2006]. We separately averaged the isolates data for countries from the Northern and the Southern hemisphere over the 1997-2006 seasons [figure 2a]. For this duration, the annual flu epidemic seems to have several peaks. The highest peak occurs in summer. It is interesting to note the different epidemic dynamics in China, with an incidence peak in the summer, which is very different from other Northern hemisphere countries. China probably acts as a reservoir for the flu virus and also as a resource for transmission of the virus from pigs to humans in the summer. Thus, the summer peak in China is likely related to the role played by this virus reservoir. For each country for each year, we aligned the week with the highest peak, $week_{peak}$, to $week_1$ so that we could match the burst times for different years. This results in all data from a week preceding the peak week being shifted to ($week_{52}$ – $week_{peak}$), and all data from a week following the peak week being shifted to ($week_i$ – $week_{peak}$). After rearranging the data in such manner, we compared the results for average isolates from our simulation with those from the WHO FluNet data.

*Multiple-component vaccine*

We evaluated the viability of using a multiple-component vaccine against simultaneous introduction of multiple viral strains. In our model, we considered the possible scenario where two initial viruses, differing by $p_{epitope}$, cause a pandemic. A $p_{epitope}$ value of zero indicates that viruses are the same, whereas a $p_{epitope}$ value of 1 indicates that the viruses are completely different. The model then predicted the efficacy of both single and multiple-component vaccines as a function of the $p_{epitope}$ (figure 4a). The cumulative attack rates are also predicted (figure 4b). A multiple-component vaccine was found to exhibit greater efficacy for the case where multiple-strains are introduced. There always exists a lag between viral outbreaks and administration of pertinent vaccine. We considered the effect such lag has on single and multiple-component vaccines. It was found that the lag affects single-component vaccine more than it affects multiple-component vaccine (figure 4c and 4d).

*Population at risk*

In order to manage the risk of possible introduction of multiple strains of avian influenza, we combined together our stochastic model and a bioinformatics study. Such risk management methods are in common use in financial institutions [Hull, 2005]. In economics, Value at Risk (VaR) is used as a metric in risk management studies. The term stands for the maximum loss over a given period of time with a certain confidence level. We use an analogous term, Population at Risk (PaR) in our study. It defines the maximum percentage of the world's population that is at risk from viral infection in the event of an influenza pandemic over the period of one year with a confidence level. As an example, a PaR value of 0.1 at the 95% confidence level means that if there are 100 pandemics, the maximum percentage of the global population that would be infected by the circulating viral strains in 95 of these pandemics stands at 0.1. The method could be applied to the risk management of infectious diseases in general. In the case of avian influenza, we analyzed all H5N1 strains in the National Center for Biotechnology Information (NCBI) database [NCBI database, 2007], and calculated the $p_{epitope}$ for all pairs of the viral strains. The average value turned out to be $p_{epitope} = 0.118$. We generated two initial viruses using the $p_{epitope}$ value, and carried out simulations for 2000 yearly pandemics for 3 cases: with no vaccination, vaccination with single-component vaccine, and vaccination with multiple-component vaccine. The PaR was calculated for confidence interval between 0.9 and 0.99. As seen from figure 5, our model is able to predict the virus transmission (figure 5a) and basic production number (figure 5b). It also shows that the fewer the vaccine components, the more the number of people that get infected in these pandemics (figure 5c). This clearly shows that a multiple-component vaccine not only improves the efficacy of a vaccine, but is also important to manage worse case scenarios.

*Summary*

We have developed a pandemic model for avian influenza. Virus evolution, inter-personal transmission, vaccination, and immune history are taken into account. The model is able to reproduce the observed data for H3N2 isolates. Vaccination against multiple-components is not a foregone conclusion in case of simultaneous introduction of multiple strains. The parameter $p_{epitope}$, helps decide if use of multiple-component vaccine is likely to be beneficial. The $p_{epitope}$ measure of antigenic distance also helps decide what strains to incorporate in the vaccine under such scheme. The model shows that a vaccine with multiple-components outperforms a single-component vaccine when there are greater than one circulating strains that are antigenically different from each other. The model also shows that compared to the traditional single-component vaccine, a multiple-component vaccine is affected to a lesser degree by the lag between viral outbreak and vaccine administration. This result has important implications if there is a sudden attack of avian influenza pandemic, and the process of designing, producing, delivering, and administrating a vaccine is delayed significantly. The model proves that the spread of the virus is contingent upon number of emergent strains, fraction of the population vaccinated, and the time of administration of the vaccine following the outbreak. Lastly, the model provides a new parameter, PaR, that foretells how severe a potential epidemic could be. Combined with multiple-component vaccine, PaR would be valuable to quantify the benefit from actual vaccination strategies, beyond the broad notion of addressing multiple isolates. One can calculate virus differences through WHO FluNet to get an average $p_{epitope}$ value that could be used in our model. The model can yield PaR values for different areas of the world. The results of PaR calculation could alert the policy makers and general public of the current pandemic risk level, and along with $p_{epitope}$ calculations also help pharmaceutical companies to prepare optimal vaccine components. An improved predictive ability regarding the extent of impending epidemic and appropriate vaccine design would enhance our ability to better manage precious public health resources.

**Appendix:**

*Sensitivity analysis*

We carried out a sensitivity analysis of the model for all parameters for figure 5. A sensitivity analysis is important for model verification. It also indicates to which public health policy measures an influenza outbreak may be most sensitive. First, we did sensitivity analysis for the network structure. The variable parameters were the exponents in the power law relation of population distribution ($N_{city}$), group distribution ($N_{group\_contact}$) and flight distribution ($N_{city\_contact}$). For the population distribution, when the exponent was decreased by 5%, the cumulative attack rate was reduced by 41%, and when the exponent was increased by 5%, the cumulative attack rate was increased by 62%. For the group distribution, when the exponent was decreased by 4%, the cumulative attack rate was increased by 27%, and it decreased by 19% for an increase of the same magnitude in the variable. The epidemic curve remained practically unaffected for changes in both these variables. For the flight distribution, a 2.5% decrease in the exponent caused a jump of nearly 85%, and an increase of 2.5% in the same variable caused a decrease of 50% in the cumulative attack rate. In the epidemic trend, one can see the average isolations decrease almost 50% at week 40, for an increase of 2.5% in the exponent value.

We checked the model's sensitivity against the flight transportation parameters. A change of 7% in Ffac in either direction resulted in a proportional change of 28% in the cumulative attack rate. The epidemic trend curve was practically unchanged, with only a small jump in average isolations around week 40 with increasing Ffac. A change of 8% in either direction for ffactor1 also resulted in proportional change of 7% for the cumulative attack rate. The epidemic curve remained unaltered for these changes. When FCap was decreased by 15%, the cumulative attack rate was decreased by 43%, and when FCap was increased by 15%, the rate jumped by 65%. The cumulative attack rate has a non-linear relationship with the FCap. The epidemic curve was affected by these changes in FCap. Around week 10, and week 40, the average cases increased up to two fold, when FCap increased from 100 to 115. Reducing air travel would seem to be effective for mitigating an influenza epidemic.

Sensitivity analysis was carried out for the initial virus transmission probability, $\tau_0$, the immune killing, cKill, and for the immune history capacity. For a 7% change in $\tau_0$, the cumulative attack rate changed by nearly 40%. The function increased linearly with the variable. The epidemic trend saw almost two-fold jump at week 40 in the average cases. The cumulative attack rate increased nearly 39% when cKill decreased by 6%. It decreased by 32% for an increase of same magnitude in cKill. The epidemic trend showed nearly 50% reductions in average cases at week 40, for a 6% increase in cKill. The cumulative attack rate remained independent of the immune history capacity. A two fold increase in the parameter brought about 1% decrease in the attack rate. The epidemic trend remained unaltered. Measures such as face masks would seem to be effective.

Next, we checked the dependence of the cumulative attack rate on the mutation rate, $R_{mutation}$, and the transmission probability of the mutants, cProb. A two-fold increase in $R_{mutation}$ rendered the cumulative attack rate the same (<1% change). A two-fold decrease in the variable yielded the same result. The epidemic trend saw nearly 3-fold increase in the average isolations around week 40 when $R_{mutation}$ was doubled. The cProb parameter, when decreased by 50%, saw a decrease of 1% in the cumulative attack rate, but when the parameter was increased two-fold, the cumulative attack rate jumped by nearly 10%. This indicated a non-linear relationship between the function and the variable. The epidemic trend showed nearly 3 times increase in the average cases at week 40 when the parameter was doubled.

We changed the initially infected group number. The sensitivity analysis showed that doubling the number of initially infected groups pushed the cumulative attack rate up by 22%, whereas halving the value of the parameter decreased the attack rate by nearly 13%. The epidemic trend curve remained unchanged for these variations.

The fixation rate depends up on the mutation rate, $R_{mutation}$, and the mutation transmission probability, cProb, the most. We carried out a sensitivity analysis against these two variables. For both the parameters for non-dominant epitopes, the fixation rate stayed relatively flat. However, for the dominant epitope, the fixation rate nearly doubled with a two-fold increase in the variable.

We carried out sensitivity tests for single-component vaccine at $p_{epitope} = 0.0$ and $p_{epitope} = 1.0$, and for multiple-component vaccine at $p_{epitope} = 1.0$, both at day 40 with 40% of the population vaccinated.

The network structure parameter (exponent for population distribution) affects cumulative attack rate and efficacy. The cumulative attack rate increases while efficacy decreases with increasing value of the parameter. For the case of $p_{epitope} = 1.0$, a change of 5% in the parameter results in 2 to 4 fold change in the cumulative attack rate for both single and multiple-component vaccines. A similar trend is seen for the case of $p_{epitope} = 0.0$ for single-component vaccine. The efficacy drops by 6 to 7% for multiple-component vaccine at $p_{epitope} = 1.0$ as well as for single-component vaccine at $p_{epitope} = 0.0$, and it drops by around 10% for the single-component vaccine at $p_{epitope} = 1.0$ for every 5% increase in the value of the network structure parameter.

We checked the sensitivity of cumulative attack rate and efficacy against exponent of flight distribution. Once again, in terms of absolute values, a multiple-component vaccine outperforms a single-component vaccine over the tested range of the parameter. The cumulative attack rate decreases with increasing value of the variable. For the case of $p_{epitope} = 1.0$, a change of 4% in the parameter results in 40% to 50% change in the cumulative attack rate for both single and multiple-component vaccines. A similar trend is observed for the case of $p_{epitope} = 0.0$ for a single-component vaccine. For $p_{epitope} = 1.0$, and for the same change in the parameter, the efficacy of single-component vaccine increases around 4%, and that of multiple-component vaccine increases by about 5%, whereas, for $p_{epitope} = 0.0$, the efficacy for the single-component vaccine increases by 5%.

Sensitivity analysis of the flight transportation parameter gives the following results. In the case that $p_{epitope} = 1.0$, when we change Ffac by 7%, the cumulative attack rate changes by around 40% for the multiple-component vaccine, and it changes by 45% for the single-component vaccine. The efficacy drops by 6% for the multiple-component case, and it drops by nearly 10% for the single-component vaccine. When $p_{epitope} = 0.0$, the cumulative attack rate of the single-component vaccine changes by nearly 50%, and its efficacy changes by nearly 5% for a 7% change in Ffac. The attack rate increases, whereas, efficacy decreases with increase in the value of Ffac.

With FCap as the variable, the sensitivity analysis shows that for the case of $p_{epitope} = 1.0$, with a 15% increase, the cumulative attack rate more than doubles, and efficacy drops by 15% for the multiple-component vaccine, and the cumulative attack rate doubles, and efficacy drops by nearly 13% for the single-component vaccine. For the case of $p_{epitope} = 0.0$, there is a two-fold increase, and a 15% decrease for cumulative attack rate and efficacy, respectively.

We analyzed the sensitivity of our model to the virus transmission probability, $\tau_0$. For the case of $p_{epitope} = 1.0$, and both for multiple as well as single-epitope vaccine, the cumulative attack rate changes nearly 50% for a 7% change in the value of $\tau_0$. The efficacy varies 6% and 12% respectively for the two cases. For the case of $p_{epitope} = 0.0$, the single-component vaccine sees cumulative attack rate change 50% and efficacy change 6% for a 7% change in the value of $\tau_0$. The attack rate increases, and efficacy decreases with increasing $\tau_0$.

For the case of $p_{epitope} = 1.0$, a change of 6% in intrinsic killing probability, cKill, changes the cumulative attack rate by 3 fold, and 2.5 fold respectively, for multiple and single-component vaccines. Efficacy changes by 14% in multiple-component vaccine, and by 18% in single-component vaccine. When $p_{epitope} = 0.0$, the cumulative attack rate changes 3 fold, and efficacy changes 14% for single-component vaccine for a similar change in cKill. Once again, attack rate increases, and efficiency decreases with increase in cKill.

Sensitivity analysis was also applied to the mutation rate, $R_{mutation}$. With $p_{epitope} = 1.0$, cumulative attack rate changes by 15% in multiple-component, and by 10% in single-component vaccine. For multiple-component vaccine, a drop in $R_{mutation}$ by 50% increased the efficacy by 7%, but a two-fold increase in $R_{mutation}$ decreased the efficacy by 13%. A similar trend was observed for a single-component vaccine. For the case of $p_{epitope} = 0.0$, a 50% decrease in $R_{mutation}$ raised the cumulative attack rate by nearly 25%, and reduced efficacy by nearly 3%. A two-fold increase in $R_{mutation}$ on the other hand raised the attack rate by 25%, but reduced efficacy by nearly 16%.

Finally, sensitivity analysis of the mutant transmission probability, cProb, yielded the following results. When $p_{epitope} = 1.0$, a two-fold increase in cProb brings 17% and 10% increase in the cumulative attack rate of multiple and single-component vaccines respectively. For the same change in cProb, efficacies drop by 7%, and 4% for multiple and single-component vaccines respectively. When $p_{epitope} = 0.0$, for a two-fold change in cProb, the cumulative attack rate of the single-component vaccine shows a jump of around 25%. The efficacy drops by around 7% for the same change in cProb.

**Acknowledgements**
This research was supported in part by the FunBio program of DARPA.

Table 1: Summary of results for $p_{epitope}$ Analysis [Gupta et. al., 2006]; Individual references also provided below.

| Year | Vaccine strain | Circulating strain | Vaccine efficacy(%) | Dominant epitope | $p_{epitope}$ | $p_{sequence}$ | $d_1$ | $d_2$ | $N_u$ | $N_v$ |
|---|---|---|---|---|---|---|---|---|---|---|
| 1971-1972 | Aichi/2/68(V01085) | HongKong/1/68(AF201874) | 7 [Smith et. al., 1979] | A | 0.158 | 0.033 | | | 25202 | 26317 |
| 1972-1973 | Aichi/2/68(V01085) | England/42/72(AF201875) | 15 [ Smith et. al., 1979] | B | 0.190 | 0.055 | | | 26130 | 26778 |
| 1973-1974 | England/42/72(ISDNENG72) | PortChalmers/1/73(AF092062) | 11 [ Smith et. al., 1979] | B | 0.143 | 0.018 | 5 [Smith et. al., 1999] | 4 [Smith et. al., 1999] | 26536 | 28158 |
| 1975-1976 | PortChalmers/1/73(AF092062) | Victoria/3/75(ISDNVIC75) | -3 [ Smith et. al., 1979] | B | 0.190 | 0.055 | 4 [Kendal et. al., 1983] | 16 [Kendal et. al., 1983] | 25591 | 29247 |
| 1984-1985 | Philippines/2/82(AF233691) | Mississippi/1/85(AF008893) | -6 [Keitel et. al., 1988] | B | 0.190 | 0.033 | 2 [WHO, 1988] | 2 [ WHO, 1988] | 241 | 171 |
| 1985-1986 | Philippines/2/82(AF233691) | Mississippi/1/85(AF008893) | -2 [Keitel et. al, 1997; Demicheli et.al., 2004] | B | 0.190 | 0.033 | 2 [ WHO, 1988] | 2 [ WHO, 1988] | 25391 | 15388 |
| 1987-1988 | Leningrad/360/86(AF008903) | Shanghai/11/87(AF008886) | 17 [Edwards et. al., 1994; Keitel et. al., 1997] | B | 0.143 | 0.024 | 2 [ WHO, 1988] | 1 [ WHO, 1988] | 1451064 | 1211060 |
| 1989-1990 | Shanghai/11/87(AF008886) | England/427/88(AF204238) | -5 [Edwards et. al., 1994] | A | 0.105 | 0.021 | | | 1016 | 1016 |
| 1992-1993 | Beining/32/92(Af008812) | Beining/32/92(Af008812) | 59 [Campbell et. al., 1997] | | 0.0 | 0.0 | 0 [Ellis et. al., 1995] | 0 [Ellis et. al., 1995] | 131 | 131 |
| 1993-1994 | Beining/32/92(Af008812) | Beining/32/92(Af008812) | 38 [Demicheli et. al., 2004] | | 0.0 | 0.0 | 0 [Ellis et. al., 1995] | 0 [Ellis et. al., 1995] | 12 | 26 |
| 1994-1995 | Shangdong/9/93(Z46417) | Johannesburg/33/94(AF008774) | 25 [Nichol et. al., 1995] | A | 0.108 | 0.021 | | | 424 | 422 |
| 1995-1996 | Johannesburg/33/94(AF008774) | Johannesburg/33/94(AF008774) | 45 [Grotto et. al., 1998] | | 0.0 | 0.0 | 0 [CDC, 1997; Corias et. al., 2001] | 0 [CDC, 1997; Corias et. al., 2001] | 652 | 684 |
| 1996-1997 | Nanchang/933/95(AF008725) | Wuhan/359/95(AF008722) | 28 [Millot et. al., 2002] | B | 0.095 | 0.006 | 0 [CDC, 1997; Corias et. al., 2001] | 0 [CDC, 1997; Corias et. al., 2001] | 2978 | 273 |
| 1997 | Nanchang/933/95(AF008725) | Wuhan/359/95(AF008722) | 11 [Mixeu et. al., 2002] | B | 0.095 | 0.006 | 0 [CDC, 1997; Corias et. al., 2001] | 0 [CDC, 1997; Corias et. al., 2001] | 299 | 294 |
| 1997-1998 | Nanchang/933/95(AF008725) | Sydney/5/97(AJ311466) | -17 [Bridges et. al., 2000] | B | 0.238 | 0.040 | 4.5 [Corias et. al., 2001; Pontoriero et. al., 2001] | 27.3 [Corias et. al., 2001; Pontoriero et. al., 2001] | 554 | 576 |
| 1998-1999 | Sydney/5/97(AJ311466) | Sydney/5/97(AJ311466) | 34 [Bridges et. al., 2000] | | 0.0 | 0.0 | 0 [Corias et. al., 2001; Cox wt. al., 2003] | 0 [Corias et. al., 2001; Cox wt. al., 2003] | 596 | 582 |
| 1999-2000 | Sydney/5/97(AJ311466) | Sydney/5/97(AJ311466) | 43 [Lester et. al., 2003] | | 0.0 | 0.0 | 0 [Corias et. al., 2001; Cox wt. | 0 [Corias et. al., 2001; Cox wt. | 324 | 342 |

| | | | | | | | al., 2003] | al., 2003] | | |
|---|---|---|---|---|---|---|---|---|---|---|
| 2001-2003 | Panama/2007/99(ISD NCDA001) | Panama/2007/99(ISD NCDA001) | 55 [Kawai et. al, 2003] | | 0.0 | 0.0 | 0 [Corias et. al., 2001; Cox wt. al., 2003] | 0 [Corias et. al., 2001; Cox wt. al., 2003] | 982 | 968 |
| 2003-2004 | Panama/2007/99(ISD NCDA001) | Fujian/411/2003(ISDN 38157) | 12 [Dolan et. al., 2004] | B | 0.143 | 0.040 | 2 [Cox et. al., 2003] | 8 [Cox et. al., 2003] | 402 | 1000 |

The definitions for $p_{epitope}$, $p_{sequence}$ are described previously (fractional change in the immunodominant epitope and fractional change in the whole sequence respectively). The terms $d_1$ and $d_2$ are the two available measures of distance based on ferret anti-serum. Figures below are to provide the analysis of the data presented in the table. The bracketed numbers are the references used. When more than one antisera assay has been performed, the calculated distances are averaged. Error bars are calculated assuming binomial statistics for each data set: $\epsilon^2 = [\sigma_v^2/u^2/N_v + (v/u^2)^2 \sigma_u^2/N_u]$, where $\sigma_v^2 = v(1-v)$ and $\sigma_u^2 = u(1-u)$ If two sets of data are averaged in 1 year, then $\epsilon^2 = \epsilon_1^2/4 + \epsilon_2^2/4$.

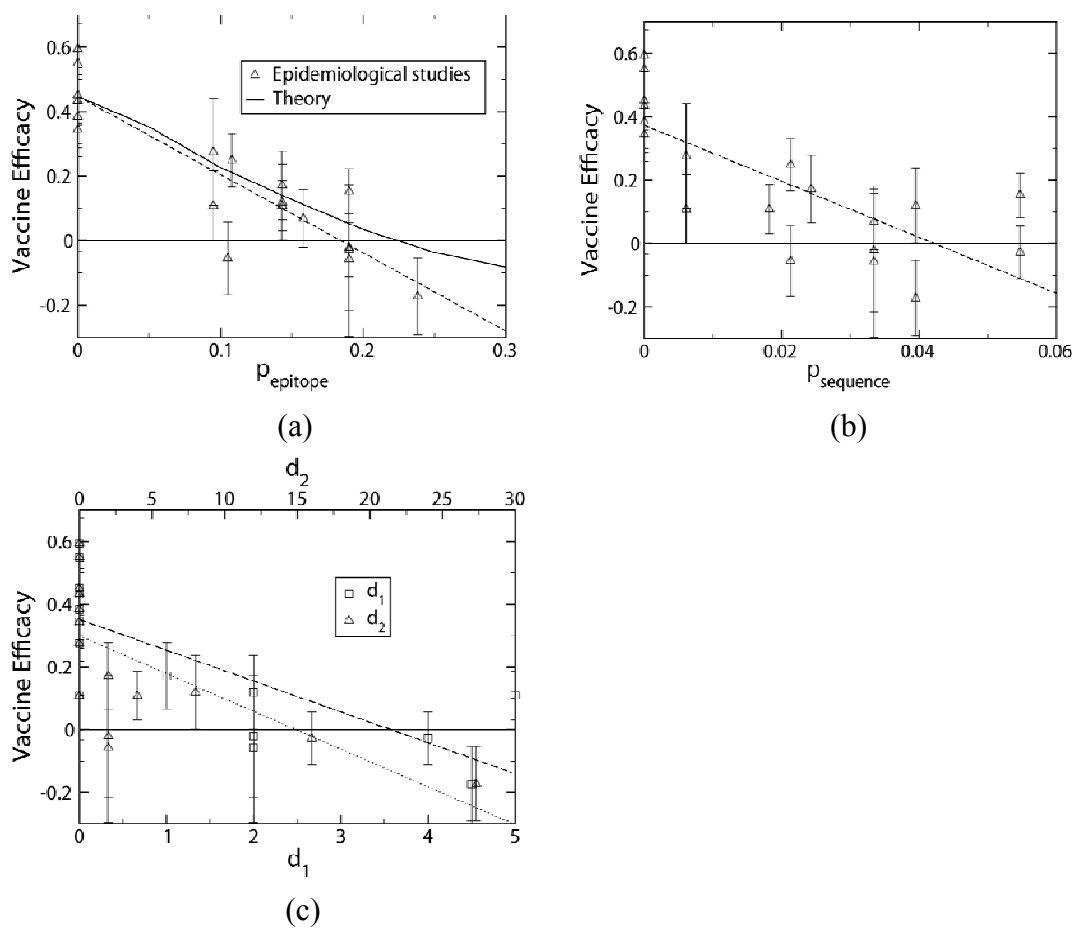

Figure 1: Vaccine efficacy for influenza-like illness as observed in epidemiological studies as a function of three different measures of antigenic distance (Gupta et. al., 2006). A linear least squares fit to the data is also shown. Error bars are one standard error ε, calculated as described in Table 1. (a) $p_{epitope}$: (long dashed, $R^2$=0.81). (b) $p_{sequence}$: (long dashed, $R^2$=0.59). Figure shows the same epidemiological data as in (a). Only the definition of the x-axis is different. (c) $d_1$ (long dashed, $R^2$=0.57) and $d_2$ (short dashed, $R^2$=0.43), derived from ferret antisera experiments. Results were averaged when multiple hemagglutinin inhibition (HI) studies had been performed for a given year. These HI binding arrays measure the ability of ferret antisera to block the agglutination of red blood cells by influenza viruses. The epidemiological data is same as in figure (a). Only the definition of the x-axis is different.

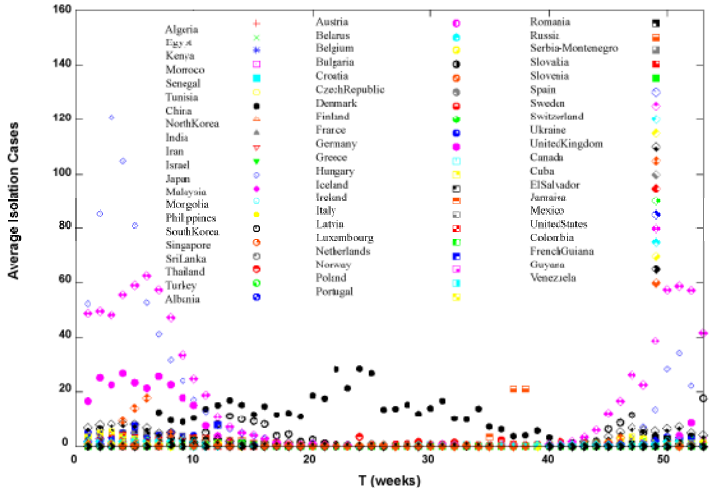

(a)

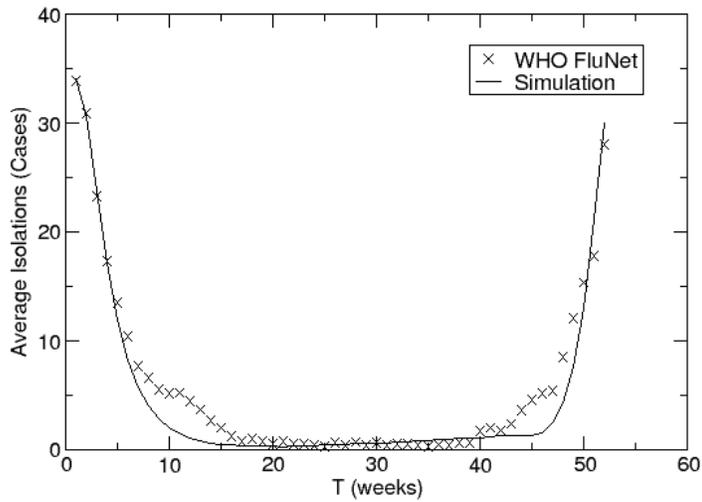

(b)

Figure 2: Average isolates cases of human H3N2 influenza. (a) Cases reported during last 10 years for various countries around the world. (b) Comparison between the WHO FluNet database and as predicted by our immunological and epidemiological model.

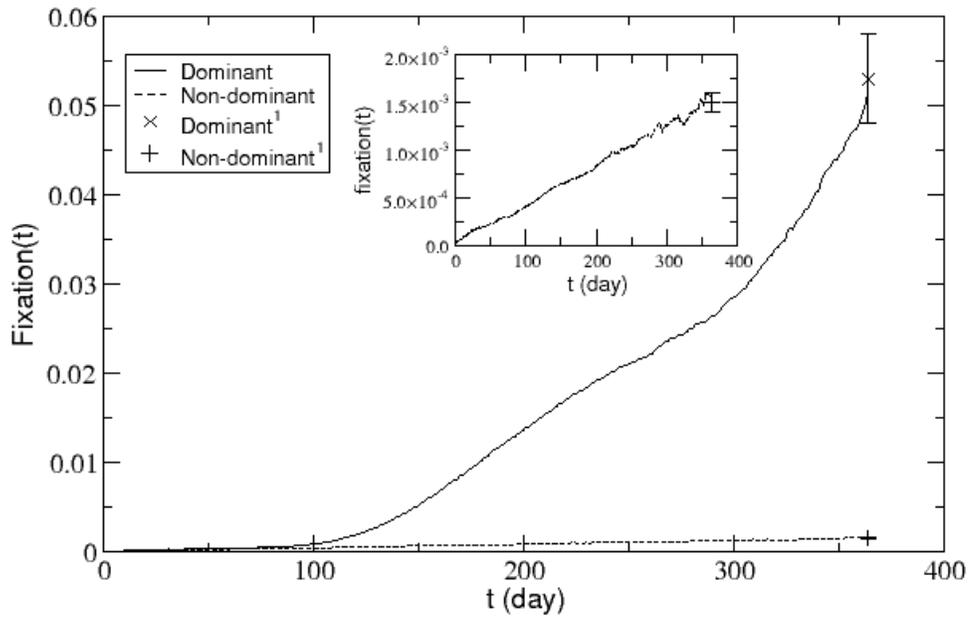

Figure 3: Influenza fixation rate for dominant and non-dominant epitopes. Dominant: fixation rate for dominant epitope from simulation, Non-dominant: fixation rate for non-dominant epitopes from simulation; Dominant[1]: fixation rate and error bar for dominant epitopes from sequence analysis [Ferguson et. al., 2003], Non-dominant[1]: fixation rate and error bar for non-dominant epitope from sequence analysis [Ferguson et. al., 2003].

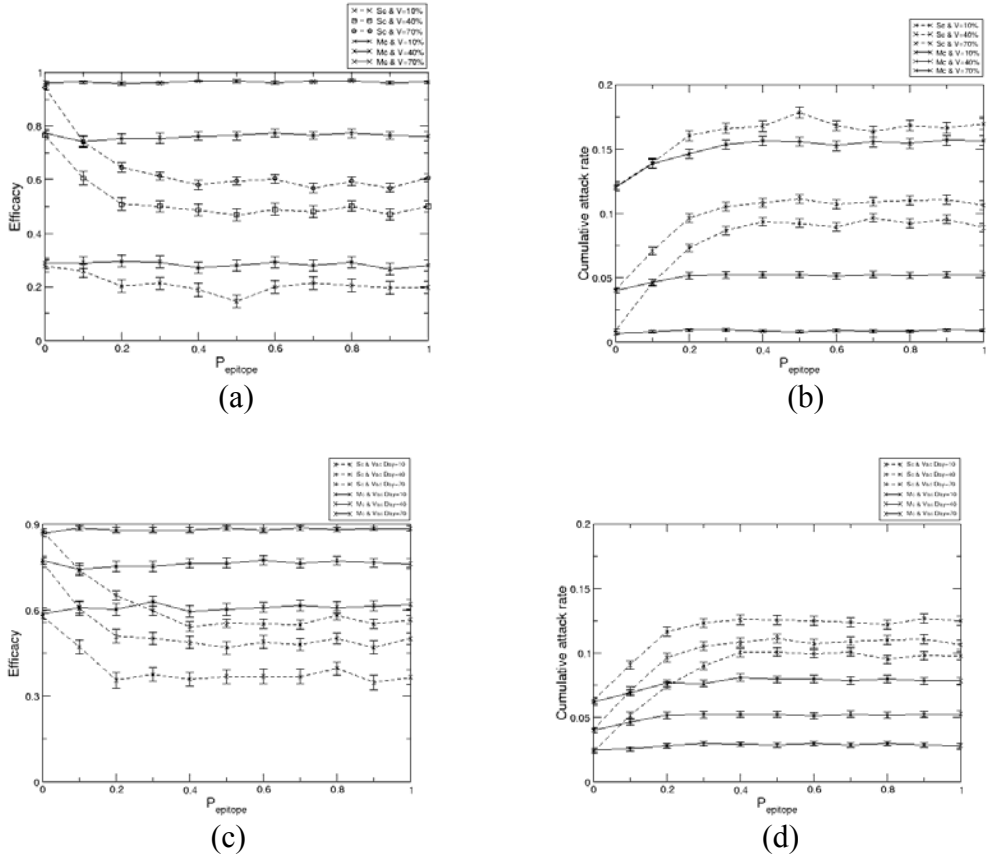

Figure 4: The effect of a vaccine for the initial introduction of two-strains. This indicates the multiple-component vaccine excels single-component vaccine for different vaccination day and vaccination populations. Sc: single-component vaccine. Mc: multiple-component vaccine. V: vaccination population over total population. Vac Day: The day when vaccine is administered. Vaccination population dependency at Vac Day = 40: (a) Efficacy as a function of $p_{epitope}$ for single and multiple component vaccines with different vaccination populations. (b) Cumulative attack rate as a function of $p_{epitope}$. Vaccination day dependence at V=40%: (c) Efficacy as a function of $p_{epitope}$. (d) Cumulative attack rate as a function of $p_{epitope}$.

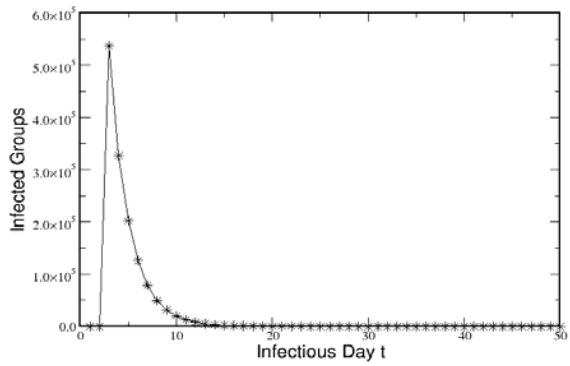

(a)

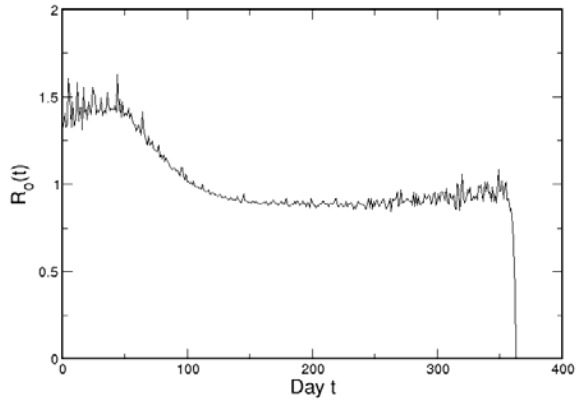

(b)

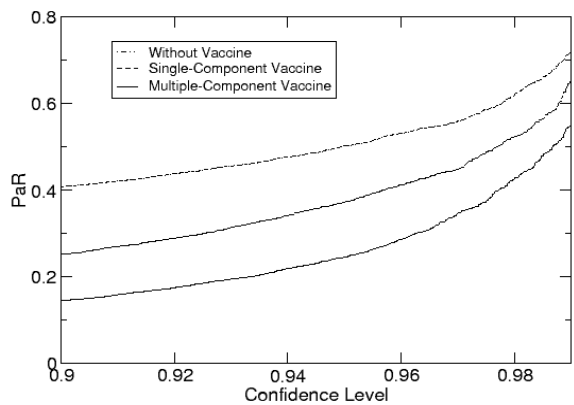

(c)

Figure 5: Virus transmission, basic production number, and Population at Risk (Par) from our immunological and epidemiological model. (a) Averaged infectious cases. (b) The basic production number ($R_0$). (c) Population at Risk (PaR) over a year for three cases: without vaccination, single-component vaccination, and multiple-component vaccination.